\documentclass{article}
\usepackage{spconf,amsmath,graphicx,hyperref,amssymb,xcolor,colortbl,booktabs, epstopdf, caption, enumitem}
\DeclareCaptionFont{my9pt}{\fontsize{9pt}{11pt}\selectfont}

\title{Speech-to-See: End-to-End Speech-Driven Open-Set Object Detection}
\name{ Wenhuan Lu \textsuperscript{1}, Xinyue Song \textsuperscript{1}, Wenjun Ke \textsuperscript{2}, Zhizhi Yu \textsuperscript{1*}, Wenhao Yang \textsuperscript{1}, Jianguo Wei \textsuperscript{1}
\thanks{*Corresponding author}
}
\address{
\textsuperscript{1} College of Intelligence and Computing, Tianjin University, Tianjin, China \\
 \textsuperscript{2} PipeChina Institute of Science and Technology, Tianjin, China
}

\begin{document}
%
\maketitle

\begin{abstract}
Audio grounding, or speech-driven open-set object detection, aims to localize and identify objects directly from speech, enabling generalization beyond predefined categories. This task is crucial for applications like human-robot interaction where textual input is impractical. However, progress in this domain faces a fundamental bottleneck from the scarcity of large-scale, paired audio-image data, and is further constrained by previous methods that rely on indirect, text-mediated pipelines. In this paper, we introduce Speech-to-See (\textit{Speech2See}), an end-to-end approach built on a pre-training and fine-tuning paradigm. Specifically, in the pre-training stage, we design a Query-Guided Semantic Aggregation module that employs learnable queries to condense redundant speech embeddings into compact semantic representations. During fine-tuning, we incorporate a parameter-efficient Mixture-of-LoRA-Experts (MoLE) architecture to achieve deeper and more nuanced cross-modal adaptation. Extensive experiments show that \textit{Speech2See} achieves robust and adaptable performance across multiple benchmarks, demonstrating its strong generalization ability and broad applicability.
\end{abstract}
\section{Introduction}
\label{sec:intro}
\begin{figure*}[t!]
    \centering
    \includegraphics[width=\textwidth]{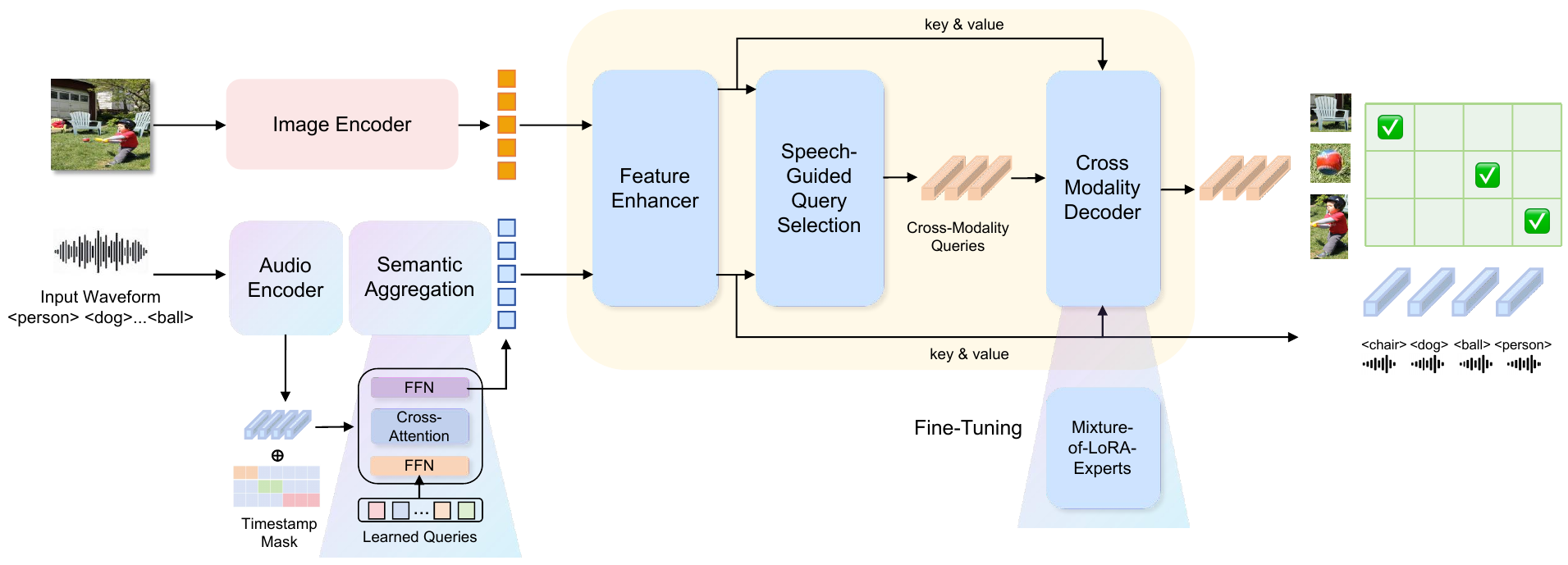}
    \caption{Our proposed end-to-end audio grounding architecture: \textit{Speech2See}.}
    \label{fig:framework}
\end{figure*}
Open-set object detection extends closed-set detection by recognizing not only known but also novel categories~\cite{GROUNDING-DINO}. Motivated by the way humans rely on language for semantic association when interpreting visual information, recent research has explored multimodal learning to construct shared visual-language representations from large-scale image–text pairs~\cite{ALIGN,clip}. This approach enables broader category coverage and zero-shot detection. Representative approaches include GLIP~\cite{GLIP}, YOLO-World~\cite{YOLO-WORLD}, and Grounding DINO~\cite{GROUNDING-DINO}, all of which rely on text to provide semantic guidance for object localization and identification.  

However, most existing studies remain limited to image–text modalities~\cite{OV-DETR,REGIN-CLIP,DET-CLIP}. In real-world applications like human–robot interaction and voice-guided navigation, text input is often unavailable. Speech, as a natural and intuitive communication channel, carries acoustic cues beyond text~\cite{human-robot2}, making speech-driven open-set detection a promising yet underexplored direction.

A pioneering attempt is YOSS~\cite{yoss}, which adopts a two-stage design combining pre-trained audio and vision models. It first aligns an audio encoder to visual features via textual mediation through CLIP~\cite{clip}, and then integrates the aligned audio encoder into a detection framework~\cite{yolov8} using contrastive learning to map speech to image regions. However, this pipeline suffers from two limitations. First, the decoupled two-stage training hinders efficient end-to-end optimization, leading to suboptimal convergence~\cite{ROD-MLLM}. Second, its reliance on text mediation limits direct cross-modal fusion, underutilizing the rich acoustic cues and properties of speech~\cite{vosoughi2025can}.
Moreover, the scarcity of audio–image pairs poses a fundamental bottleneck for progress in audio grounding research.



To address the aforementioned limitations, we propose a novel end-to-end method, Speech-to-See (\textit{Speech2See}), by leveraging semantic knowledge from large-scale pre-trained models. Our approach leverages two key sources of knowledge: the rich text-image mapping semantics from Grounding DINO~\cite{GROUNDING-DINO} and the deep acoustic representations from unsupervised audio encoders such as HuBERT~\cite{hubert}. 
It is noteworthy that a modality gap exists between these two types of knowledge, necessitating modality alignment. Therefore, \textit{Speech2See} employs a progressive pre-training and fine-tuning paradigm that efficiently bridges the audio and visual modalities, thereby efficiently leveraging transferred prior knowledge.

During pre-training, we introduce a lightweight Query-Guided Semantic Aggregation (QSA) module. Acting as a semantic adapter, QSA distills raw speech features into a compact representation that can be directly fused with visual features. During fine-tuning, although the transferred decoder parameters retain semantic knowledge from text–image alignment, they remain suboptimal for speech-guided decoding. To deepen the alignment, we enhance the decoder with a Mixture-of-LoRA-Experts (MoLE) architectures, inspired by LLaVA-MoLE~\cite{lava-moe}. Extensive experiments demonstrate that \textit{Speech2See} achieves state-of-the-art performance in audio grounding, effectively overcoming the challenges of data scarcity and the limitations of prior methods.

\section{Methodology}
\label{sec:method}

\subsection{Overview}

As illustrated in Fig.~\ref{fig:framework}, \textit{Speech2See} employs a novel progressive pre-training and fine-tuning paradigm to efficiently transfer rich semantic knowledge from a foundational, pre-trained grounding method to the speech domain.

In pre-training, given an image–audio pair (I,A), we firstly extract multi-scale visual features via a Swin Transformer~\cite{swin_transformer} and raw speech embeddings via HuBERT~\cite{hubert}. Our Query-based Semantic Aggregation (QSA) module then condenses the redundant HuBERT embeddings into compact semantic tokens. These tokens and the visual features are subsequently processed within a Grounding DINO-inspired~\cite{GROUNDING-DINO} architecture, which employs a feature enhancer for deep fusion and a speech-guided query selection module to initialize object queries. Finally, a cross modality decoder integrates these enhanced features to generate predictions, thereby establishing a foundational alignment between speech and vision.
During fine-tuning, a parameter-efficient Mixture-of-LoRA-Experts (MoLE) is incorporated into the decoder to refine this alignment. The overall architecture thus establishes direct speech–vision correspondence, supports efficient end-to-end optimization, and exhibits strong generalization.

\subsection{Query-based Semantic Aggregation}
The audio encoder~\cite{hubert} produces redundant feature sequences that are not well-suited for direct alignment with visual modalities. Therefore, we introduce a set of $K$ learnable queries $Q = {q_k \in \mathbb{R}^d}_{k=1}^K$ where $K \ll N_t$, acting as a semantic adapter to condense features into compact tokens.

Given audio embeddings $X_A=\{x_j \in \mathbb{R}^d\}_{j=1}^{N_t}$, we employ a Transformer-based cross-attention mechanism where each query attends to the entire sequence. The aggregated token of query $q_i$ is:
\begin{equation}
O_i = \mathrm{Attn}(q_i, X_A) = 
\sum_{j=1}^{N_t} 
\mathrm{softmax}\!\left(\frac{q_i \cdot x_j}{\sqrt{d}}\right) \cdot x_j,
\end{equation}
yielding $O=\{O_i\}_{i=1}^K$ as the final compact tokens.  
This query-based aggregation efficiently transforms raw acoustic data into a high-level semantic form, and aggregated tokens can subsequently be leveraged within the cross modality decoder, enabling more effective multimodal fusion.

\subsection{Parameter-Efficient Cross-Modal Adaptation}
\begin{figure}[t]
    \centering
    \includegraphics[width=0.8\linewidth]{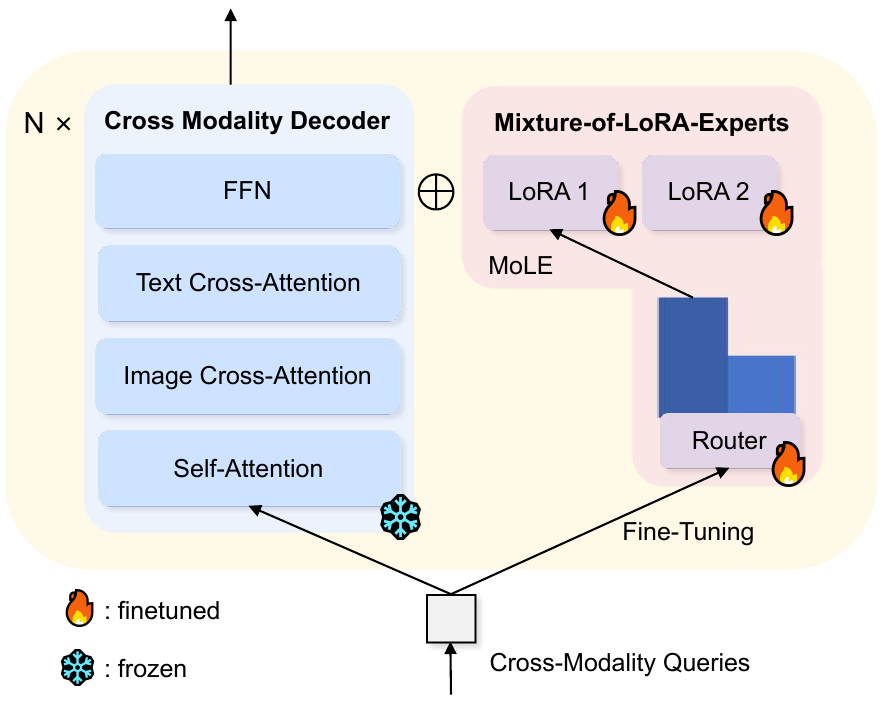}
    \caption{MoLE plugin integrated into the cross-modal decoder.}
    \label{fig:moe}
\end{figure}

The transferred decoder parameters inherently encapsulate text–image semantic knowledge. Although initial alignment is performed through Query-based Semantic Aggregation, these parameters remain suboptimal for speech modality decoding. To achieve deeper alignment, we integrate a Mixture-of-LoRA-Experts (MoLE)  architecture (Fig.~\ref{fig:moe}), inspired by LLaVA-MoLE~\cite{lava-moe}, into the decoder during fine-tuning.
This mechanism enables a more intricate and context-aware integration of modalities.
Specifically, each layer is equipped with a router and $K$ LoRA~\cite{lora} experts $\{E_1, \dots, E_K\}$.  
Given an input query $q$, the router computes expert scores and selects the most relevant expert via Top-1 routing:
\begin{equation}
k^* = \arg \max_{j=1 \dots K} G_j(q) = \arg \max_{j=1 \dots K} (W_j^g q),
\end{equation}
where $W_j^g$ denotes the learnable routing weight of the $j$-th expert. Then the selected expert $E_{k^*}$ is activated to perform computation, and the feed-forward output can be updated as:
\begin{equation}
f_{\mathrm{FFN}}^{\prime}(q) = f_{\mathrm{FFN}}(q) + E_{k^*}(q),
\end{equation}
where each expert adopts the standard LoRA formulation with low-rank decomposition:
\begin{equation}
E_{k^*}(q) = \frac{\alpha}{r} B_{k^*} A_{k^*} q.
\end{equation}
Each linear transformation in the feed-forward network is equipped with a dedicated MoE, while sharing a common router across the layer.  

\subsection{Training Objective}
The overall training loss consists of two components, including the \textit{detection loss} and the \textit{load balancing loss} for MoLE.

\textbf{Detection Loss.}  
The detection loss $\mathcal{L}_{\text{det}}$ combines three terms, namely an L1 loss and a GIoU loss~\cite{giou} for bounding box regression, and a contrastive loss adapted from GLIP~\cite{GLIP} for classifying predicted objects based on speech tokens.

\textbf{Load Balancing Loss.}  
Following prior work on sparse MoE~\cite{lava-moe, switch-transformer}, a load balancing loss $\mathcal{L}_{lb}$ is introduced in each MoLE layer to prevent expert underutilization.  

Finally, the training objective is defined as:
\begin{equation}
    \mathcal{L}_{\text{total}} = \mathcal{L}_{\text{det}} + \alpha \cdot \mathcal{L}_{lb},
\end{equation}
where $\alpha$ is a hyperparameter scaling the \textit{load balancing loss}.
\section{Experiments and Results}
\label{sec:typestyle}

\subsection{Implementation Details}

\noindent\textbf{Datasets.}
We conduct experiments on three widely used multimodal benchmarks following the standard open-set object detection protocol.  
Since these datasets do not provide speech annotations, we construct a multi-speaker audio corpus by converting textual descriptions into spoken utterances using \texttt{edge-TTS}\footnote{\url{https://github.com/rany2/edge-tts}}.  
To improve diversity and robustness, we generate speech with 10 distinct speakers and introduce random variations in timbre during synthesis.  
The overall dataset statistics are summarized in Table~\ref{tab:Multi-Modal Datasets}.

\begin{table}[!ht]
    \centering
    \caption{Multi-Modal Datasets with Synthesized Audio.} 
    \label{tab:Multi-Modal Datasets} 
    \resizebox{\linewidth}{!}{
    \begin{tabular}{l | c c c c c}
    \toprule
        Dataset & Type & Categories & Images & Audios & BBox\\
    \midrule
        COCO 2017~\cite{coco} & Detection & 80 & 330K & 80 & 1.5M\\
        Objects365~\cite{obj365} & Detection & 365 & 609K & 365 & 10M\\
        Flickr30k~\cite{flickr} & Grounding & -- & 31K & 158,915 & --\\
    \bottomrule
    \end{tabular}}
\end{table}

\noindent \textbf{Configuration.}
Our method is built upon Grounding DINO~\cite{GROUNDING-DINO}, with Swin-T~\cite{swin_transformer} as the visual backbone and HuBERT-Base~\cite{hubert} as the speech backbone.  
The detector is configured with 900 queries, and the speech input is downsampled to 256 tokens.  
Similar to Grounding DINO~\cite{GROUNDING-DINO}, the detection loss combines a contrastive loss, an L1 bounding box regression loss, and a GIoU loss~\cite{giou}. 
During Hungarian matching, the weights are set to $2.0/5.0/2.0$ and later adjusted to $1.0/5.0/2.0$.  
For Mixture-of-LoRA-Experts fine-tuning, a load-balancing loss is applied to each layer, averaged across layers, and scaled by $\alpha = 1 \times 10^{-2}$.

\noindent \textbf{Training Strategy.} 
During the \textit{pre-training stage}, the Swin-T, HuBERT backbones and decoder are frozen, while all other modules are optimized using AdamW with a learning rate of $1 \times 10^{-4}$, a batch size of 64, and 10 training epochs with a learning rate warm-up schedule.  
In the \textit{fine-tuning stage}, a MoLE module with rank 64 is inserted into the feed-forward network, where each insertion contains two LoRA experts.  
Only the MoLE parameters are updated for 2 epochs, while all other parameters remain frozen.  
All experiments are conducted on 8$\times$ NVIDIA A800 GPUs.

\subsection{Experimental Results}
We evaluate the performance of \textit{Speech2See} under both closed-set and zero-shot detection settings, and further examine the contribution of the Mixture-of-LoRA-Experts (MoLE) fine-tuning strategy.  

\textbf{Closed-set Audio-Visual Detection.}  
We conduct a closed-set evaluation on the COCO2017-val benchmark, as shown in Table~\ref{tab:performance_coco}. Compared with the two-stage approach of YOSS, our approach achieves performance gains. This validates our end-to-end design, which establishes a direct speech-vision alignment that avoids the information bottlenecks and error propagation inherent in multi-stage systems.

\textbf{Zero-shot Detection on COCO.}  
We evaluate the zero-shot performance of \textit{Speech2See} on the COCO benchmark. Since YOSS is trained with COCO data, it fails to satisfy the zero-shot protocol. Therefore, we only compare with text-image baselines. As shown in Table~\ref{tab:performance_coco}, \textit{Speech2See} achieves competitive results. Remarkably, \textit{Speech2See} even surpasses YOSS’s closed-set performance. These results affirm the vital role of text-image prior knowledge and progressive modality alignment in achieving robust speech-driven object detection.  Nevertheless, a performance gap remains compared to text-driven models, reflecting the greater complexity of speech understanding and its sensitivity to noise and speaker variation.

\begin{table}[!ht]
    \centering
    \caption{Results on COCO Detection Benchmarks.}
    \resizebox{\columnwidth}{!}{
        \setlength{\tabcolsep}{5pt}
    \begin{tabular}{l c c c c c}
        \toprule
        Model & Pre-Training Data & MoLE & AP & AP50 & AP75 \\
        \midrule
        \multicolumn{6}{c}{Close-Set Setting} \\
        \midrule
        YOSS-base & Flickr, COCO & $\times$ & 34.0 & 47.2 & 36.9 \\ 
        YOSS-large & +GQA & $\times$ & 39.2 & 53.3 & 42.6 \\ 
        \rowcolor{gray!20} Ours & COCO & $\times$ & 54.1 & 70.2 & 59.6 \\ 
        \rowcolor{gray!20} Ours & COCO & $\checkmark$ & 56.2 & 71.3 & 60.7 \\
        \midrule
        \multicolumn{6}{c}{Zero-Shot Setting} \\
        \midrule
        Grounding-DINO-T & O365 & $\times$ & 46.7 & - & - \\ 
        Grounding-DINO-T & O365, Flickr, GQA & $\times$ & 48.1 & - & - \\ 
        \rowcolor{gray!20} Ours & Obj365 & $\times$ & 38.2 & 49.9 & 40.8 \\ 
        \rowcolor{gray!20} Ours & Obj365 & $\checkmark$ & 39.8 & 50.3 & 41.9 \\ 
        \rowcolor{gray!20} Ours & O365, Flickr, GQA & $\times$ & 40.4 & 52.7 & 44.2 \\ 
        \rowcolor{gray!20} Ours & O365, Flickr, GQA  & $\checkmark$ & 42.7 & 55.7 & 46.8 \\
        \bottomrule
    \end{tabular}
    }
    \label{tab:performance_coco}
\end{table}

\begin{table}[!ht]
    \centering
    \caption{Results on LVIS Zero-shot Detection Benchmarks.}
    \resizebox{\columnwidth}{!}{
        \setlength{\tabcolsep}{4pt}
    \begin{tabular}{l c c c c c c}
        \toprule
        Model & Pre-Training Data & MoLE & AP & APr & APc & APf \\
        \midrule
        Grounding-DINO-T & O365, Flickr, GQA & $\times$ & 25.6 & 14.4 & 19.6 & 32.2 \\
        YOSS-base & Flickr, COCO & $\times$ & 13.6 & 4.4 & 9.5 & 18.8 \\
        YOSS-large & +GQA & $\times$ & 16.3 & 6.3 & 9.2 & 16.9 \\
        \rowcolor{gray!20} Ours & O365, Flickr, GQA & $\times$ & 18.5 & 7.6 & 11.2 & 18.2 \\
        \rowcolor{gray!20} Ours & O365, Flickr, GQA & $\checkmark$ & 19.9 & 7.9 & 13.3 & 18.9 \\
        \bottomrule
    \end{tabular}
    }
    \label{tab:performance_lvis}
\end{table}
\textbf{Zero-shot Detection on LVIS.}  
We further evaluate \textit{Speech2See} on the long-tailed LVIS benchmark, which contains 1,203 categories. As shown in Table~\ref{tab:performance_lvis}, this task remains highly challenging for speech-based models. Notably, our approach achieves a performance improvement over the two-stage YOSS model across all metrics. Despite these improvements, a gap still exists compared with text-driven detectors, reflecting the inherent complexity of speech understanding.

\textbf{Effect of Mixture-of-LoRA-Experts Fine-tuning.}  
The results in Table~\ref{tab:performance_coco} and Table~\ref{tab:performance_lvis} demonstrate that incorporating the MoLE architecture consistently improves performance in both closed-set and zero-shot settings. This validates our hypothesis that a single adaptation is insufficient for the complexities of real-world speech. By adaptively routing inputs to specialized experts, MoLE refines the initial alignment, preventing the transferred knowledge from being diluted by speech's inherent diversity and paralinguistic cues. This allows the model to learn more nuanced cross-modal mappings, leading to more robust and accurate detection.

\subsection{Visualization}
\begin{figure}[t!]
\centering
\begin{minipage}[b]{.95\linewidth}
  \centering
\centerline{\includegraphics[width=\linewidth]{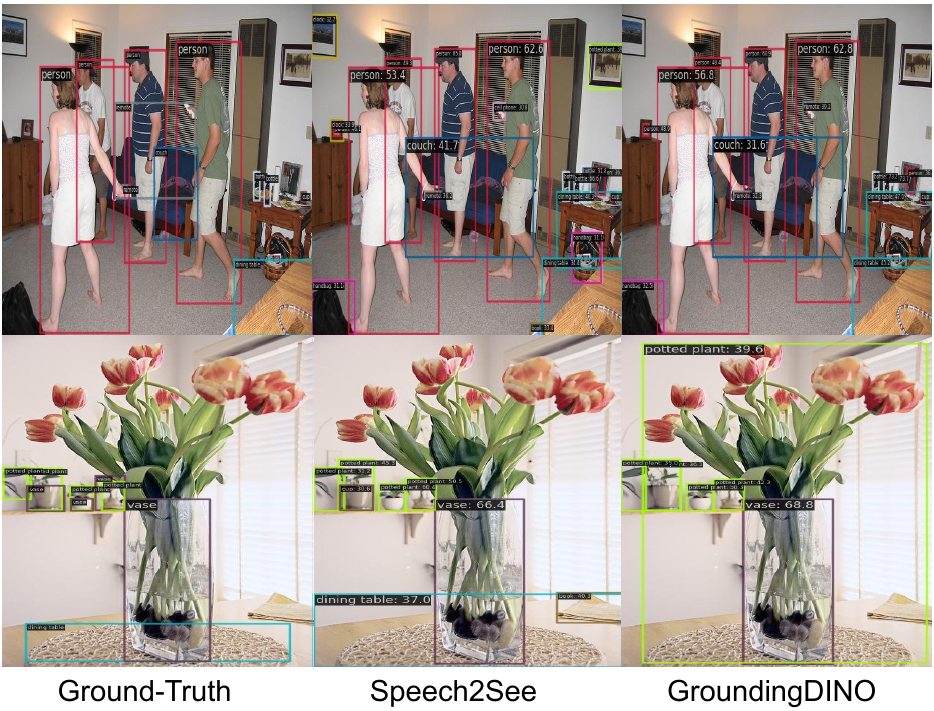}}
\end{minipage}
\caption{Qualitative Comparison of Object Detection Results}
\label{fig:visual}
\end{figure}

Fig.~\ref{fig:visual} provides a qualitative comparison among ground-truth annotations (left), predictions from \textit{Speech2See} (middle), and the text-driven baseline Grounding DINO (right). The examples illustrate both the strengths and limitations of each model. In the first row, our model exhibits robust detection: it not only matches the annotated objects but also identifies an additional item, the ``handbag'', which was omitted from the ground truth. The second row demonstrates a more nuanced case: \textit{Speech2See} achieves an accurate localization of the ``dining table'' compared to the ground truth, whereas Grounding DINO fails to detect the object altogether. Nonetheless, our model overlooks the ``potted plant''. These results suggest that while \textit{Speech2See} effectively transfers knowledge to support accurate speech-driven localization, both speech and text based detectors exhibit complementary strengths and weaknesses that merit deeper analysis.

\subsection{Ablation Studies}
We compare the computational efficiency of our end-to-end \textit{Speech2See} with a cascaded baseline, which consists of an ASR module (Whisper~\cite{whisper}) followed by a text-guided detector (Grounding DINO~\cite{GROUNDING-DINO}). As shown in Table~\ref{tab:rtf}, our end-to-end architecture is more efficient in terms of parameter count and real-time factor (RTF), providing a significant advantage in processing speed. These results underscore the computational benefits of adopting an end-to-end modeling paradigm.

\begin{table}[!ht]
    \centering
    \small
    \caption{Cascaded vs. End-to-End Architecture Comparison.}
    \resizebox{\columnwidth}{!}{
        \setlength{\tabcolsep}{10pt}
    \begin{tabular}{cccc}
    \toprule
        Method & Backbone & Params & RTF \\ 
        \midrule
        Cascaded & Hubert + Grounding DINO & 266.7M & 0.41 \\ 
        End-to-End  & Speech2See & 197.8M & 0.35 \\ 
    \bottomrule
    \end{tabular}}
    \label{tab:rtf}
\end{table}


\section{Conclusion}
\label{sec:conclusion}
This paper proposes Speech-to-See (\textit{Speech2See}), an end-to-end framework for audio grounding. Our approach addresses two key challenges, data scarcity and limitation of previous methods, by transferring knowledge and establishing speech-vision alignment through a progressive training paradigm. A Query-based Semantic Aggregation module is introduced to generate compact representations for audio inputs.
The subsequent fine-tuning stage leverages a Mixture-of-LoRA-Experts architecture to provide parameter-efficient and adaptive refinement for diverse speech inputs. Experiments on multiple benchmarks demonstrate consistent improvements over existing methods in both closed-set and zero-shot settings. Future work will focus on addressing long-tailed distributions and improving robustness to noisy and diverse speech.

\vfill\pagebreak

{\small
\bibliographystyle{IEEEbib}
\bibliography{refs}
}





\end{document}